# Do Some Virtual Bound States Carry Torsion Trace?

Richard James Petti, ORCID: 0000-0002-9066-7751
146 Gray Street, Arlington, Massachusetts 02476 U.S.A. E-mail: rjpetti@gmail.com

17 February 2022

**Abstract**. This article presents theoretical arguments that certain virtual bound states carry the trace component of affine torsion. The motivation for this work is that Einstein–Cartan theory, which extends general relativity by including torsion to model intrinsic angular momentum, is becoming more credible. The author is not aware of any situation for which there is evidence or substantial argument for the presence of torsion trace, except in the continuum theory of edge dislocations in crystals. The main evidence for the hypothesis consists of analogies between the structure of virtual bound states and (a) geometry of dislocations in crystal lattices, which are modeled with torsion; and (b) modeling of intrinsic angular momentum by torsion in Einstein–Cartan theory and the theory of micro-elasticity. The work focuses on conjectured presence of torsion in para-positronium, which intermediates annihilation of an electron and a positron with opposite z-spins. If the virtual bound state carries torsion, then the local law of conservation of angular momentum can hold over the spacelike separation during annihilation.

## Contents

## 1. Introduction

This article presents theoretical arguments to support the hypothesis that certain virtual bound states carry the trace component of affine torsion. The motivation for this work is that Einstein–Cartan theory (EC), which extends general relativity by including torsion to model intrinsic angular momentum (a.m.), is becoming more credible [Blagojević and Hehl 2013; Petti 2021; Trautman 2006]. The author is not aware of any situation for which there is evidence or substantial argument for the presence of torsion trace, except in the continuum theory of edge dislocations in crystals. The evidence presented here is of two types that match the structure of virtual bound states: (a) angular momentum (a.m.) and conservation thereof are represented as torsion in EC and in continuum theories of exchange of intrinsic and orbital a.m.; (b) screw and edge dislocations in crystal lattices, which are represented as torsion, illustrate the geometry of torsion.

We define "intrinsic" angular momentum (a.m.) to be a.m. on too small a scale to be modeled as orbital a.m. in a particular classical model; that is, intrinsic a.m. cannot be modeled by displacement fields (in solids) or velocity fields (in fluids). Intrinsic a.m. includes but is not limited to quantum spin.

## 1.1. Torsion and angular momentum

This article contains enough information about the role of affine torsion in physics to support the main hypothesis of this work. This work is not a comprehensive review of torsion in physics. It discusses affine torsion as a classical tensor field define in E. Cartan's general theory of affine connections [Cartan 1922, 1923, 1924, 1925; Bishop and Crittenden 1964; Kobayashi and Nomizu 1963, 1969; Tu 2018] and as identified with intrinsic a.m. in EC.

Below are main events in the use of torsion to represent intrinsic a.m.

a) Early in the 20[th] century, E. Cosserat pioneered the field of micro-elasticity in continuum mechanics, which allows points to have torque and classical intrinsic a.m. Cosserat showed that exchange of intrinsic and orbital a.m. requires an intrinsic a.m. tensor and a non-symmetric momentum tensor during the exchange. This is the earliest example of modeling intrinsic a.m. with affine torsion of which the author is aware [Cosserat 1909].

b) In the 1920s, E. Cartan developed the theory of affine connections, in which Riemannian curvature (rotational curvature) and affine torsion (translational curvature) are natural companions in Riemann–Cartan affine geometry. Several times in the 1920s, Cartan urged Einstein to extend general relativity (GR) to include torsion. In 1929 Einstein replied that he did not understand Cartan's description of torsion, and he did not know what physical process torsion would model [Debever 1979].

c) In the early 1960s, EC was essentially completed with the interpretation of EC as a gauge theory of the Poincaré group, and by identifying torsion as intrinsic a.m. [Kibble 1961; Sciama 1962]. EC is a modest extension of GR that enables conservation of a.m. to include exchange of intrinsic and orbital a.m., which GR cannot do because of its symmetric momentum tensor [Blagojević and Hehl 2013].

Torsion has applications in EC that are important in gravitational theory but that are not relevant to our current purpose. (a) Repulsive torsion force interacts with ultra-high squared density of intrinsic a.m. to eliminate gravitational singularities. (b) Current cosmological models ignore intrinsic a.m. that comes from two sources: fluid turbulence [Monin A S, Yaglom 1971; Peshkov et al 2019a; 2019b], and exchange of intrinsic and orbital a.m. across scales from galactic clusters ($\sim 10^{23}$ m) to orbits of moons ($\sim 10^9$ m).

Capozziello et al. discuss several methods of decomposition of torsion into irreducible representations, and alternative applications of torsion, including theories in which torsion is defined as the gradient of a scalar field [Capozziello 2001].

Jiminez et al. assert that three classes of gravitational gauge theories are acceptable: (a) GR, (b) flat space theories based on torsion, and (c) flat space theories based on nonmetricity [Jiminez et al 2013].

According to [Blagojevich and Hehl 2013], only three known theories satisfy all the empirical tests of gravitational theories: GR, EC, and teleparallel gravity, which uses torsion in a flat space.

The author prefers EC, for four reasons:

a) EC makes the least invasive changes to GR that enable exchange of intrinsic and orbital a.m. In the absence of intrinsic a.m., EC is identical to GR.

b) Torsion appears in EC exactly where it appears in Cartan's master structure equations for affine differential geometry; EC has no ad-hoc torsion terms.

c) EC can be derived from GR with no additional assumptions [Petti 2021].
d) EC has been more thoroughly studied than the alternatives.

### 1.2. Torsion and geometry of lattice defects

Physical theories that use affine torsion to model continuum distributions of defects in affine lattices provide insight into the geometry of torsion. Since the 1950s, the continuum theory of dislocations models distributions of dislocations and disclinations with torsion and Riemannian curvature respectively. [Bilby 1957; Kondo 1955]. The discrete theory of dislocations provides the best intuitive geometrical model of discrete torsion.

### 1.3. Results

This article uses insights from EC and the theory of dislocations to support the conjecture that certain virtual bound states carry torsion trace that enables micro-scale conservation of a.m. along spacelike separations. This work focuses on annihilation of two spin ½ particles with opposite z-spins that are connected over a spacelike separation by a virtual bound state, for example annihilation of an electron and a positron with opposite z-spins, mediated by para-positronium.

The annihilation of two Dirac particles joined at the ends of their world lines by a virtual bound state has these analogs in the fields discussed here.

- In the discrete theory of dislocations, the configuration that is analogous to para-positronium is two screw dislocations whose dislocation lines end, and the ends are joined by an edge dislocation.
- In the continuum theory of dislocations, the analogous configuration is two densities of antisymmetric torsion (screw dislocations) of opposite helicities, such that, wherever the two densities decline together, torsion trace appears so that the conservation law of dislocation lines is satisfied.
- In Einstein–Cartan theory, the analogous configuration is two densities of antisymmetric torsion (Dirac particles) of opposite helicities with spacelike separation, such that, wherever the two densities decline together, torsion trace appears so that conservation of a.m. is satisfied. In Riemann–Cartan geometry, terms quadratic in torsion appear, so that zero net torsion does not imply zero torsion.
- The law of conservation of dislocation lines and the law of conservation of total a.m. in EC are both applications of the contracted first Bianchi identity of Riemann–Cartan geometry.

Models of dislocations in terms of torsion are presented in section 2. The model of intrinsic a.m. in terms of torsion in EC is described in section 3. The model of virtual bound states in terms of holonomy and torsion is described in section 4. Speculations about future developments related to this work are discussed in section 5. The interpretations of torsion in Riemann–Cartan geometry, the theory of dislocations, and EC are compared in Table 1 in the Appendix.

## 2. Dislocations in lattices illustrate geometry of torsion

The theory of dislocations in affine lattices provides a well-accepted physical application of torsion [Kröner and Anthony 1975; Roychowdhury and Gupta 2013; Yavari and Goriely 2012]. We begin with an affine defect that is not a dislocation and is more familiar to physicists than torsion.

### 2.1. Disclinations

A disclination is a discrete unit of Riemannian curvature. Figure 1 is a graphic representation of a disclination with less than $2\pi$ radians around its vertical centerline. In a real disclination the two faces of the angular deficit are joined together. This dispiration is a discrete version of Riemannian curvature $R_{\mu\nu\mu\nu}$.

To construct a 2-dimensional version of a disclination, excise a wedge of material of angle θ from a flat disk. This yields a horizontal slice of the 3-D graphic in Figure 1. If you join the edges of the cut, the disk becomes a conical surface. The angular deficit at the apex of the cone is θ, and the area deficit of a disk centered at the apex of radius r is ½ θ r². The tip of the cone has a Dirac delta of Riemannian curvature of magnitude θ. The remainder of the cone has no intrinsic curvature.

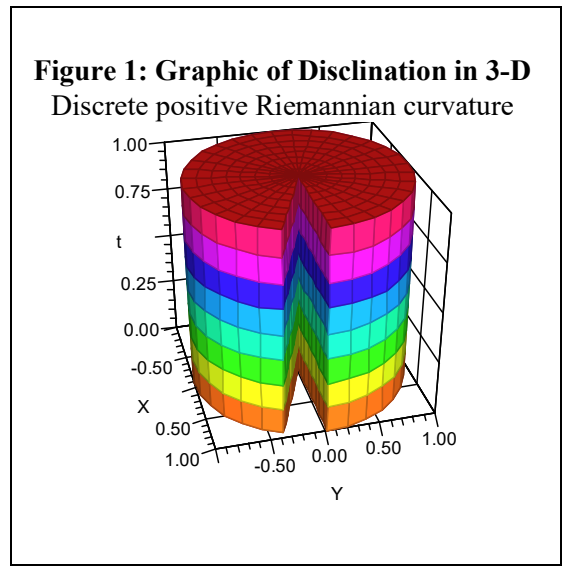

**Figure 1: Graphic of Disclination in 3-D**
Discrete positive Riemannian curvature

To construct a surface with positive Riemannian curvature, remove many small angular wedges whose centers are distributed around the disk. The limit of many small wedges yields a surface with positive curvature.

A disclination that has more than 2 π radians at the center represents discrete negative curvature. Negative Ricci curvature can occur GR in the presence of a positive cosmological constant, which introduces repulsive acceleration. A distribution of negative disclinations provides a discrete geometric model for dark energy via the cosmological constant.

### 2.2. Discrete dislocations in lattices

In crystalline materials, a dislocation exists in a neighborhood where traversing a loop that would be closed in a perfect lattice is not closed. For example, in a perfect cubic lattice, a loop with 10 lattice steps in direction +x, 10 in direction +y, 10 in direction –x, and 10 in direction –y) is closed. If the loop path surrounds a dislocation line, then the path does not close. We borrow from materials science the term "Burghers vector" to denote the "failure-to-close vector" when we traverse a loop that would close in a perfect lattice, but that surrounds a dislocation in our defected lattice.

Two types of dislocations are most common in materials: screw dislocations and edge dislocations.

2.2.1. Screw dislocations

A screw dislocation has a Burghers vector that is orthogonal to the plane of the loop around the dislocation.

**Figure 2: Graphic representations of screw dislocations and edge dislocations**

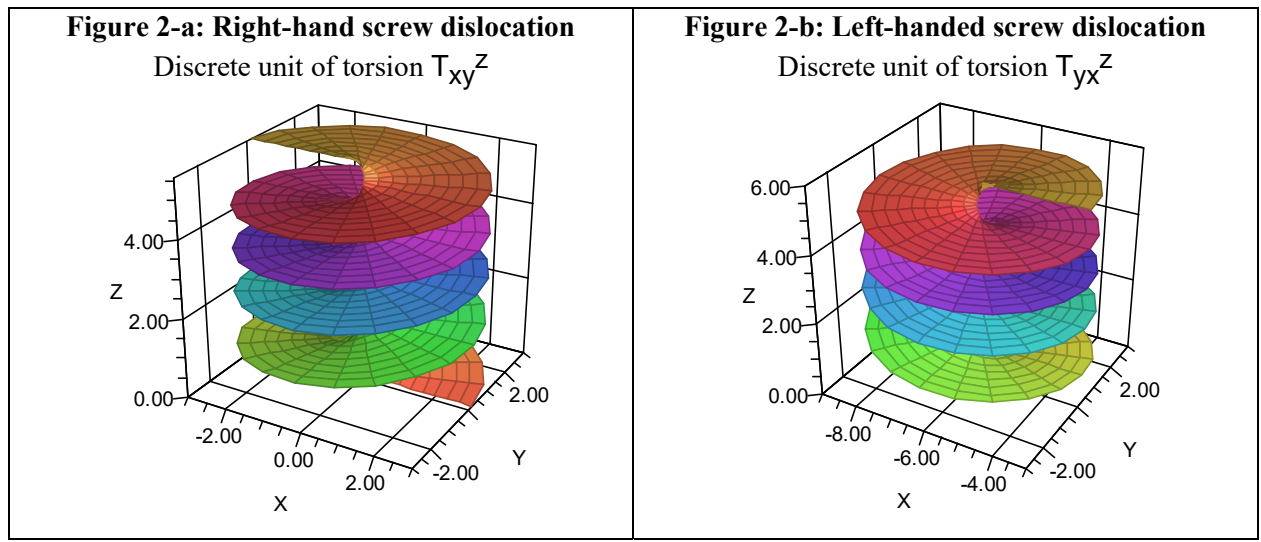

Figure 2-a: Right-hand screw dislocation — Discrete unit of torsion $T_{xy}{}^z$

Figure 2-b: Left-handed screw dislocation — Discrete unit of torsion $T_{yx}{}^z$

The graphical representation looks like a screw, or a parking garage ramp, as in Figure 2.

2.2.2. Edge dislocations

Figure 3 illustrates edge dislocations, which have Burghers vector in the plane of the loop used to measure the dislocation.

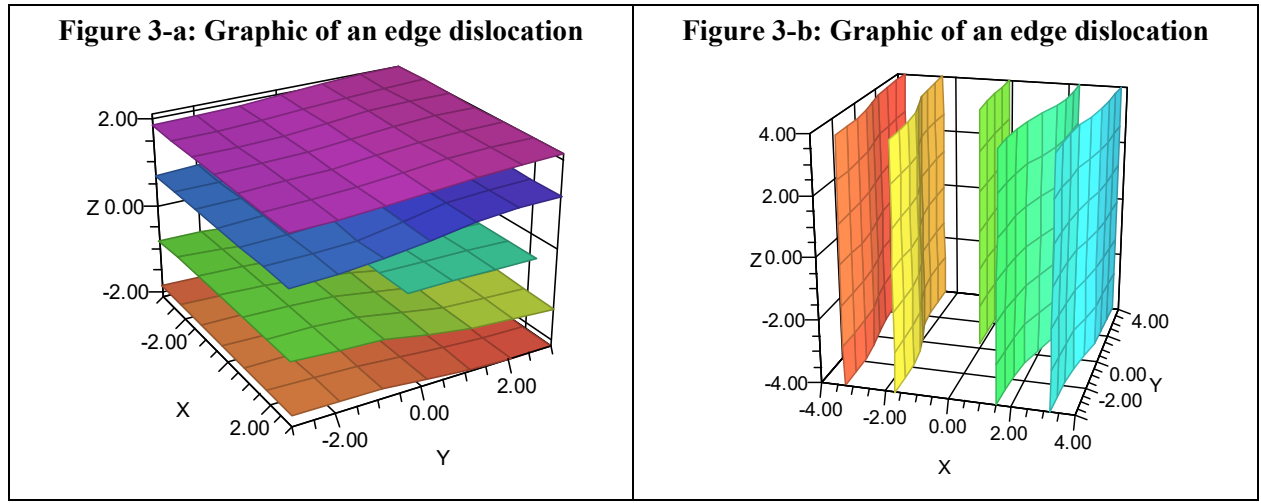

| Figure 3-a: Graphic of an edge dislocation | Figure 3-b: Graphic of an edge dislocation |

In Figure 3-a, if you travel around a loop in the z-y plane that surrounds the center, you are translated one lattice spacing in the +z direction. (Travel 4 units in direction +z, 4 units in direction +y, 4 unit in direction −z, and 4 unit in direction −y. You finish 1 unit above the point where you started.) The graph looks like a parking garage with many levels, where one level covers only half the area of the others; that half-level has an edge where you could fall off if there were no guard rail. This dislocation is a discrete unit of torsion $T_{zy}{}^z$.

Figure 3-b is a graphic representation of the same edge dislocation with different coordinate orientation. If you travel around a loop in the y-x plane that surrounds the center, you are translated one lattice spacing in the +x direction. This dislocation is a discrete unit of torsion $T_{yx}{}^x$.

2.2.3. Other defects in affine lattices

Defects in affine lattices can be classified by their dimension.

- Zero-dimensional defects (point defects): vacancies (missing atom or ion), interstitials (atoms or ions not located at a regular lattice position), Frenkel defects (displaced atoms or ions that create a void and a nearby inclusion), and substitutions (atoms or ions at regular lattice locations but not the usual kind of atoms or ions).
- One-dimensional defects: dislocations and disclinations. These are called "line defects" because the Burghers vector continues along a line that cannot end inside a crystal; this is the conservation law for dislocation lines. The lines can terminate at two-dimensional defects where the regular lattice structure breaks down.
- Two-dimensional defects: grain boundaries, phase boundaries, domain walls, stacking faults, and free surfaces.
- Three dimensional defects: precipitates and inhomogeneities.

**2.3. Continuum distribution of dislocations**

In Riemannian geometry, rotational holonomy is a discrete version of Riemannian (rotational) curvature.

(1) $$\text{rotational curvature} = \lim_{\text{area} \to 0} \frac{(\log(\text{rotational holonomy}))}{(\text{area enclosed by the loop})}$$

(Fine point: log is the inverse of the exponential map; log maps the structure group to its Lie algebra.)

In Riemann–Cartan geometry, translational holonomy is a discrete version of affine torsion.

$$\text{(2)} \qquad \text{torsion} = \lim_{\text{area} \to 0} \frac{\log(\text{translational holonomy})}{(\text{area enclosed by the loop})}$$

In the continuum theory of dislocations, torsion represents the density of dislocations. The two common types of dislocations have characteristic components of torsion.

- Screw dislocations have Burghers vectors (in the z-direction) orthogonal to the plane of the loop (the x–y plane); they are represented by torsion $T_{xy}{}^z$, or any other permutation of the orthogonal directions {x, y, z}. The antisymmetrized indices x and y indicate the plane of a loop.

- Edge dislocations have Burghers vectors (in the z direction) lying in the plane of the loop (the z–y plane); they are represented by torsion trace $T_{zy}{}^z$, or any other permutation of {x, y, z}. The antisymmetrized indices z and y indicate the plane of a loop.

## 3. Einstein–Cartan theory describes intrinsic angular momentum with torsion

### 3.1. Background

EC is the first theory of fundamental physics that uses affine torsion. EC is a gauge theory of the Poincaré group. Torsion is identified with intrinsic a.m.

- A stationary spinning particle with a.m. polarized in the xy plane (conventionally described as a.m. in the z direction) is represented as a discrete unit of torsion $T_{xy}{}^t$. Its translational holonomy (Burghers vector) is a timelike vector; parallel translation around a spacelike equatorial loop translates the observer into the future or the past.

- A stationary spin ½ Dirac field with z-spin polarized in the +z direction has a spin tensor equal to antisymmetrized torsion $T_{[x,y,t]}$. Dirac fields couple only to totally antisymmetric torsion.

Although EC is a classical theory of physics, the quantum spin fits well in EC.

### 3.2. Conservation of angular momentum

In the continuum theory of dislocations, the conservation law for dislocations is.

$$\text{(3)} \qquad D_k (T_{ij}{}^k + (\text{torsion trace terms})_{ij}{}^k) = \text{sum of rotational curvature terms}$$

where D is the covariant derivative operator and i, j, k ∈ {1,2,3} [Bishop and Crittenden 1963 p101; Kobayashi and Nomizu 1963 p 121].

The law of conservation of a.m. in EC describes the exchange of intrinsic and orbital angular momentum.

$$\text{(4)} \qquad D_k (J_{ij}{}^k) + P_{[ij]} = 0$$

where

- D is the covariant derivative operator
- $J_{ij}{}^k$ = intrinsic a.m. = $S_{ij}{}^k / \kappa$, $\kappa = (8 \pi G/c^4)$, G = universal gravitational constant
- $S_{ij}{}^k = T_{ij}{}^k + (\delta^k{}_i T_{jm}{}^m - \delta^k{}_j T_{im}{}^m)$ is the modified torsion, which includes torsion trace.
- $P_{[ij]}$ is the antisymmetric part of the linear 4-momentum. It is the local rate of loss of orbital a.m.
- i, j, λ ∈ {0,1,2,3}. The equations of EC are simplified by distinguishing spacetime indices μ, ν, λ… and current indices (or fiber indices) i, j, k… . We ignore the distinction in this article.

Equation (4) has a simple intuitive interpretation.

- The first term, the divergence of the modified torsion, is the rate at which intrinsic a.m. increases (or decreases if negative) at a spacetime location.

- The second term, the antisymmetric part of the momentum tensor, is the rate at which orbital a.m. increases (or decreases if negative) at a spacetime location.
- Equation (4) states that, since total a.m. is conserved, the sum of the changes in intrinsic and orbital a.m. must equal zero.

The law of conservation of dislocations and the law of conservation of a.m. in EC are the same identity in Riemann–Cartan geometry: the contracted first Bianchi identity. This identity in different fields has the same geometric meaning, though it has different physical interpretations.

### 3.3. Assessment of Einstein–Cartan theory

Although no empirical evidence supports EC at this time, the arguments for the theory are substantial.

a) EC and teleparallel gravity (which alters the geometric basis of GR) and GR are the only theories of gravitation known to satisfy all empirical tests of GR [Trautman 2006; Blagojević and Hehl 2013].
b) EC can model exchange of intrinsic and orbital a.m. GR cannot do this because its momentum tensor is symmetric.
    - The essence of turbulence is transfer of orbital a.m. to smaller scales [Monin and Yaglom 1971].
    - Current theories of turbulence do not track intrinsic a.m. Present research suggests that transfer of a.m. in turbulence ends with classical intrinsic a.m. [Peshkov 2019a, 2019b].
    - Classical turbulence is common in cosmology. Therefore a master classical theory of spacetime should be able to model these phenomena, at least in theory.
c) EC, plus quantum field theoretic models of matter, removes many of the singularities in GR; for example, in black holes and in some Big Bang cosmological models [Poplawski 2010a, 2010b, 2012].
d) Torsion appears in EC only in places where it occurs in Cartan's master structure equations for affine geometry. EC has no ad-hoc torsion terms.
e) Classical EC can be derived from GR without additional assumptions [Petti 2021].
f) Mathematically, EC is that it includes translational symmetries of spacetime in the structure group. So EC can properly derive linear momentum as the Noether current of spacetime translations.

These advantages arguably make EC a more credible master theory of classical spacetime physics than GR.

## 4. Annihilation of two spin ½ particles

This work focuses on a configuration of two spin ½ particles with opposite z-spins that are separated by a spatial distance, and that annihilate via a virtual bound state that connects the ends of the worldlines of the annihilating particles. For example, an electron and a positron with opposite z-spin can annihilate at a short distance by forming a virtual bound state, para-positronium, which has $S = 0$. Ortho-positronium has $S = 1$ and $M = -1, 0, 1$, so the internal a.m. content is more complicated.

Just as the torsion trace of an edge dislocation enables screw dislocations to annihilate at a distance while preserving dislocation lines, the torsion carried by para-positronium enables spin ½ particles to annihilate at a distance while locally conserving a.m.

### 4.1. Two screw dislocations whose ends are joined by an edge dislocation

Two screw dislocations whose ends are joined by an edge dislocation is a geometric analog for two spin ½ particles that annihilate at a spacelike separation mediated by a virtual bound state.

Figure 4 represents the configuration of two screw dislocations of opposite helicity whose ends are joined by an edge dislocation.

- The screw on the left is a left-handed screw dislocation depicted in Figure 2-b. Its torsion is $T_{yx}{}^z$.
- the screw on the right is a right-handed screw dislocation depicted in Figure 2-a. Its torsion is $T_{xy}{}^z$.

- The edge dislocation has the orientation depicted in Figure 3-a. Its torsion is $T_{zy}{}^z$ (no sum on indices).
- Above the screw and edge dislocations is a flat layer of affine lattice without defects, but with part of the layer cut away so that the edge dislocation is visible.

The thought experiment below helps to visualize the overall structure of this configuration of dislocations.

a) Rotate in the x-y plane around the screw dislocation on the right. This moves you up the screw, which means the Burghers vector is in the direction +z.
b) When you reach the top of the screw, turn the dislocation line (and the plane of rotation) smoothly 90 degrees to the left (in the –x direction), so that the core of the dislocation is horizontal.
c) Along the horizontal dislocation line, rotate one time around a loop in the z-y plane, using the same number of lattice lengths in the directions +z, +y, –z, –y. The path does not close, and reveals a Burghers vector in the +z direction.
d) When you reach the left end of the edge dislocation, turn the dislocation line (and the plane of rotation) smoothly 90 degrees downward (in the –z direction), so that the core of the dislocation is vertical pointing downward on the left screw.
e) Rotate around the screw dislocation on the left in the y-x plane. This moves you down the screw, which means the Burghers vector is in the direction –z.

**Figure 4: Two screws dislocations of opposite helicity joined by an edge dislocation**

- The left-handed screw dislocation on the left is depicted in Figure 2-b. Its torsion is $T_{yx}{}^z$.
- The right-handed screw on the right is depicted in Figure 2-a. Its torsion is $T_{xy}{}^z$.
- The edge dislocation has the orientation depicted in Figure 3-a. Its torsion is $T_{zy}{}^z$.
- Above the screw and edge dislocations is a flat layer of affine lattice without defects, with part of the layer removed to expose the edge dislocation.

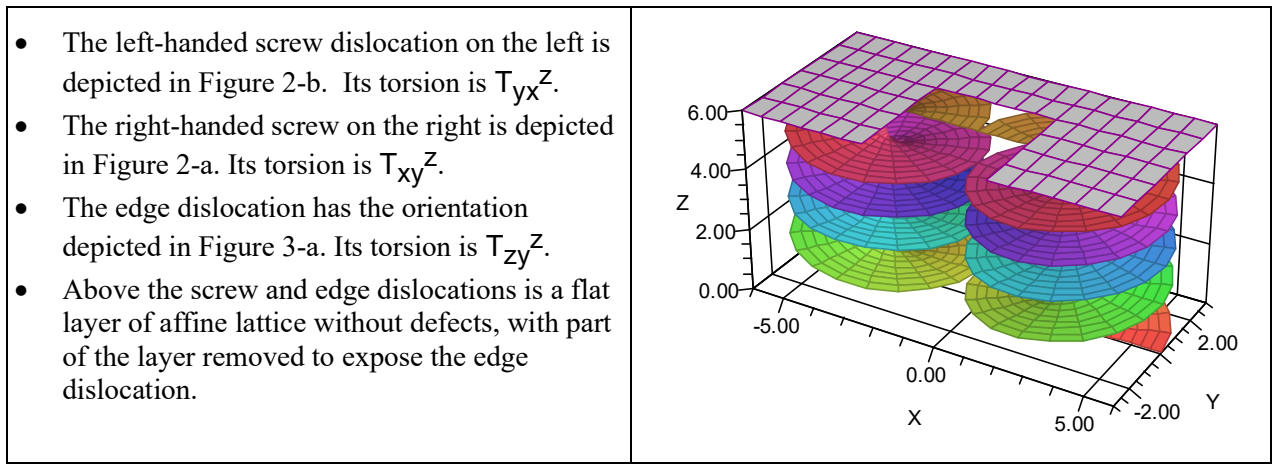

All three dislocations have Burghers vector (translational holonomy) in the +z direction. The configuration satisfies the conservation law for dislocations: dislocation lines do not end, and the Burghers vector is constant. The edge dislocation permits the screw dislocations to terminate without violating the conservation of dislocation lines.

**4.2. Two spin ½ particles of opposite z-spin joined by a virtual bound state**

The configuration of dislocations depicted in Figure 4 is a 3-D non-relativistic model for two spin ½ particles of opposite z-spin joined by a virtual bound state in relativistic quantum field theory. The dislocation model can be adapted to a relativistic model of spin ½ particles by replacing the z-direction in the dislocation model with a timelike direction in relativistic field theory. The translational holonomy (Burghers vector) in spacetime points in a timelike direction instead of the z-direction. For a Dirac particle, parallel translation once around the spiral translates to a future or past time. In Figure 4:

- The left-handed screw on the left is depicted in Figure 2-b. Its torsion is $T_{yx}{}^t$.
- The right-handed screw on the right is depicted in Figure 2-a. Its torsion is $T_{xy}{}^t$.
- The edge dislocation has the orientation depicted in Figure 3-a. Its torsion is $T_{ty}{}^t$.

The Burghers vector (angular momentum vector) lies in the time direction, as in Dirac particles. However, it is a strange form of intrinsic a.m. because its Burghers vector (a.m. vector) lies in the space × time plane of the loop used to measure it.

The author conjectures that this argument holds for any virtual bound state that mediates the annihilation of two particles with cancelling non-zero spins.

## 5. Further developments

### 5.1. Torsion in quantum gravity and other quantum theories

Intrinsic a.m. is generally more important in quantum theories than in their classical counterparts. Therefore, EC is likely a better classical starting point for a theory of quantum gravity than is GR.

Spin is omnipresent in quantum field theory. EC already assumes that all spin ½ particles have torsion. We expect that, in a future quantum field theory, torsion will be present wherever spin is present. This article is a narrow instance of how and where torsion will appear.

### 5.2. Quantum entanglement?

Quantum entanglement seems to require instant transport of information across a macroscopic spacelike separation between two entangled particles. The model proposed here suggests that certain virtual bound states use torsion trace to transport a.m. over a spacelike separation, thereby enabling local conservation of a.m. in spacetime. We might speculate that some gauge fields, analogous to torsion for spin currents, enable entangled particles to share information over macroscopic distances without violating the local structure of spacetime.

## 6. Conclusion

This work provides theoretical arguments to support the conjecture that para-positronium and possibly other the virtual bound states carry components of torsion, including torsion trace. The geometry and intrinsic a.m. in para-positronium are analogous to those of a configuration of two screw dislocations connected by an edge dislocation; that is, two pieces of antisymmetric torsion of opposite helicity connected by a spacelike line that carries torsion trace. The key properties of torsion that are used in this argument are based on analogies with two applications of torsion in classical physical theories:

- The correspondence of torsion with intrinsic a.m. is the is a central feature of EC.
- The local geometry of antisymmetric torsion and torsion trace are illustrated by screw and edge dislocations in affine lattices in crystallography.

The conservation law of total a.m. in EC and the conservation of dislocation lines in lattices have the same mathematical foundation as the contracted first Bianchi identity of Riemann–Cartan geometry.

Using torsion trace to represent a spacelike flow of intrinsic a.m. allows the local law of conservation of a.m. to hold across a spacelike separation when para-positronium mediates annihilation of two Dirac particles.

The author is unaware of any experimental tests that might confirm that these virtual bound states carry torsion. The best supports for the conjecture are (a) EC claims that all intrinsic a.m. is modeled by torsion, and (b) torsion enables conservation of total a.m. across the micro-scale spacelike separation between the fermions.

## 7. Acknowledgements

I would like to thank Professor Ilya M. Peshkov for suggesting corrections, improvements, and references in the brief discussion of turbulence; and Dr. Thomas F. Soules for suggesting improvements in organization and presentation. I thank the anonymous referee for suggesting substantial improvements in the presentation of ideas.

## Appendix: Riemann–Cartan Geometry

Riemann–Cartan geometry is the common foundation of the theory of dislocations and Einstein-Cartan theory.

**Table 1: Concepts of Riemann–Cartan geometry in dislocation theory and Einstein–Cartan theory**

| Riemann-Cartan geometry | Theory of dislocations | Einstein–Cartan theory | Quantum theory |
|---|---|---|---|
| Rotational holonomy (discrete rotational curvature) | Disclination | Continuum limit is Riemannian curvature. | Need quantum theory of gravity |
| Rotational curvature (Riemannian curvature) | Locked-in strain | Riemannian curvature (gravitation) | Need quantum theory of gravity |
| Translational holonomy (discrete translational curvature) | Burghers vector of discrete dislocation, or continuum density of dislocations | Intrinsic a.m. of a discrete particle | Conjecture: spin of discrete particle |
| Torsion (translational curvature) | density of dislocations | Density of intrinsic a.m. | Conjecture: density of intrinsic a.m. |
| Helical torsion (Burghers vector is orthogonal to plane of loop) | $T_{xy}{}^z$ = screw dislocations loop in the x-y plane, Burghers vector in z-direction. | For spin ½ fields: $T_{xy}{}^t$ = spin with loop in x-y plane and translation in t direction | Conjecture: spin density of spin ½ Dirac particle with spin polarized in z direction |
| Torsion trace | $T_{zy}{}^z$ = density of edge dislocations with loop in z-y plane, Burghers vector in z-direction. | $T_{ty}{}^t$ = density of spin with loop in the t-y plane and translation in t-direction | Conjecture: spin density of virtual bound state |
| Contracted first affine Bianchi identity | Conservation of dislocation lines. Dislocation line cannot end inside a crystal grain. | Conservation of a.m.: Δspin + Δorbital a.m. = 0 | Conjecture: conservation of a.m. |